\documentclass{article}

\usepackage{arxiv}

\usepackage[utf8]{inputenc} % allow utf-8 input
\usepackage[T1]{fontenc}    % use 8-bit T1 fonts
\usepackage{hyperref}       % hyperlinks
\usepackage{url}            % simple URL typesetting
\usepackage{booktabs}       % professional-quality tables
\usepackage{amsfonts}       % blackboard math symbols
\usepackage{nicefrac}       % compact symbols for 1/2, etc.
\usepackage{microtype}      % microtypography
\usepackage{lipsum}		% Can be removed after putting your text content
\usepackage{graphicx}
\usepackage{natbib}
\usepackage{doi}
\usepackage{amsmath}

\title{Nonlinear optical pulses in media with asymmetric gain}

\author{ \href{https://orcid.org/0000-0003-0101-3834}{\includegraphics[scale=0.06]{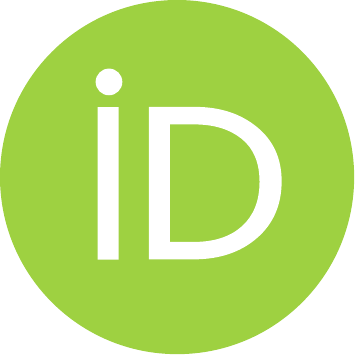}\hspace{1mm}S. K. Turitsyn} \\
	Aston Institute of Photonic Technologies\\
	Aston University\\
	Birmingham B4 7ET, UK \\
	\texttt{s.k.turitsyn@aston.ac.uk} \\
	\And
	\href{https://orcid.org/0000-0002-7820-8990}{\includegraphics[scale=0.06]{orcid.pdf}\hspace{1mm}A. E. Bednyakova} \\
	Novosibirsk State University\\
	Novosibirsk 630090, Russia \\
	\texttt{anastasia.bednyakova@gmail.com} \\
	\And
   {\hspace{1mm} E. V. Podivilov} \\
	Institute of Automatics and Electrometry SB RAS\\
	Novosibirsk 630090, Russia \\
	\texttt{podivilov@iae.nsk.su} \\
}

\renewcommand{\shorttitle}{\textit{arXiv} Template}

\hypersetup{
pdftitle={A template for the arxiv style},
pdfsubject={q-bio.NC, q-bio.QM},
pdfauthor={David S.~Hippocampus, Elias D.~Striatum},
pdfkeywords={First keyword, Second keyword, More},
}

\begin{document}
\maketitle

\begin{abstract}
	A generic novel model governing optical pulse propagation in a nonlinear dispersive amplifying medium with asymmetric (linear spectral slope) gain is introduced. We  examine the properties of asymmetric optical pulses formed in such gain-skewed media, both theoretically and numerically. We derive a dissipative optical modification of the classical shallow water equations that highlights an analogy between this phenomenon and hydrodynamic wave-breaking. We observe the development of spectral optical shock waves, and discuss the conditions and origins of this spectral wave-breaking in media with asymmetric gain. These findings provide insight into the nature of asymmetric optical pulses capable of accumulating large nonlinear phase without wave-breaking, a crucial aspect in the design of nonlinear fiber amplifiers.
\end{abstract}

% keywords can be removed

In many physical and engineering problems dealing with optical amplification a frequency dependence of the gain (that usually is broader compared to the considered signal bandwidth) is assumed to be symmetric and is often approximated by the Lorentzian spectral shape \cite{siegman1986lasers,Agrawal2010,Dienes:96}. This is, typically, a justified assumption in the spectral region near the peak of the gain curve. There are, however, relatively less explored possibilities to use edges of the gain profile that is not symmetric. 
Here we examine impact of the asymmetry in the spectral gain shape on formation and evolution of optical pulses at the carrier frequency $\omega_0$ in an amplifying nonlinear dispersive medium, considering the simplest linear asymmetric gain profile, that in the frequency domain reads:
\[ g(\omega) = g_0 - g_1 (\omega - \omega_0). \]
The slope of the gain asymmetry can be both positive or negative depending on the sign of $g_1$.  
Higher-order terms can be easily included, however, we focus here on the impact of the most general first-order approximation of the gain asymmetry. Of course, we assume that at larger deviations from $\omega_0$ linear spectral dependence will be changed and gain will not grow to infinity making the problem ill-posed. Note, that propagation of the light field affected both by gain and loss can be of interest for studies of optical systems with parity–time symmetry \cite{Ruter2010}.  

In the recent works of the Cornell group \cite{Sidorenko:19} a new type of asymmetric nonlinear pulse propagation was demonstrated,  distinguished by the presence of a dynamically-evolving gain spectrum. Nonlinear spectral broadening of the pulse led to its reshaping due to absorption and amplification. The dynamic change of the gain and spectral broadening led to  quasi-stable regimes where pulse was partially propagating at the edge of the material gain curve. We anticipate that our analysis of much more simple - "minimal" model will provide useful insight into characteristics of nonlinear pulse propagation in a medium with spectrally asymmetric amplification, beyond standard parabolic gain curve approximations.

Consider propagation of an envelope of the optical field 
$\psi(z,t)$ down the amplifying dispersive optical medium with Kerr nonlinearity within the framework of the one-dimensional generalised nonlinear Schrödinger equation (NLSE) with asymmetric gain.
\begin{equation}
	i \frac{\partial \Psi }{\partial z} - \frac{\beta_2}{2} \frac{\partial^2 \Psi}{\partial t^2} + \gamma |\psi|^2 \psi = i g_0 \Psi + g_1 \frac{\partial \Psi }{\partial t}.
    \label{eq:dimNLSE}
\end{equation}
Here $\beta_2$ is the group velocity dispersion, $\gamma$ is the nonlinear Kerr coefficient, $z$ denotes a propagation spatial coordinate, $t$ is a standard retarded time \cite{Agrawal2010} and $g_0$ and $g_1$ define a gain profile. 
The Eq.~\ref{eq:dimNLSE} is well-studied for the case $g_1=0$, when it governs parabolic pulse formation in the central energy-containing part (see e.g. \cite{PP01,PP02,PP03,PP04,Finot:06} and references therein). Self-similar parabolic pulse is an approximate  wave-breaking free solution of the Eq.~\ref{eq:dimNLSE} (with $g_1=0$). 
We would like to re-iterate that formally this equation is ill-posed due to the infinite growth of the gain at large frequencies. However, we prefer here not to introduce any formal mathematical regularisation that will change generality of the model, but instead
stress that equation is valid only in the area (in the frequency domain) around $\omega_0$ where gain can be approximated by a straight line and should not be used beyond this spectral region.

In what follows we consider only the case $g_0 =0$, because, evidently, solutions of Eq.~\ref{eq:dimNLSE} $A(z,t)$ with $g_0=0$ and $\Psi(z,t)$ with $g_0 \neq 0$ can be expressed by each other using the transform 
$A(z,t)=\Psi(z,t+\beta_2 g_0 z/g_1) \exp(i g_0 t/g_1+i 0.5 \beta_2 g_0^2 z/g_1^2),$ that has a transparent physical meaning of the effective central frequency shift by the Galilean transformation. Below, similar to \cite{Agrawal2010} we will use variable $\omega$ for the detuning around the central frequency $\omega_0$.
The Eq. \ref{eq:dimNLSE} (with $g_0=0$) has two conserved integrals, energy $E$ and the Hamiltonian $H$:
\[ E= \int{ |A(z,t)|^2 \exp[\frac{2 g_1 t}{\beta_2}] dt}\]
\[ 2 H= \int{( \beta_2 |A_{t}|^2 + 2g_1^2|A|^2/\beta_2 + \gamma |A|^4 )\exp[\frac{2 g_1 t}{\beta_2}] dt} \]

\begin{figure}[t!]
\centering\includegraphics[width=1\textwidth]{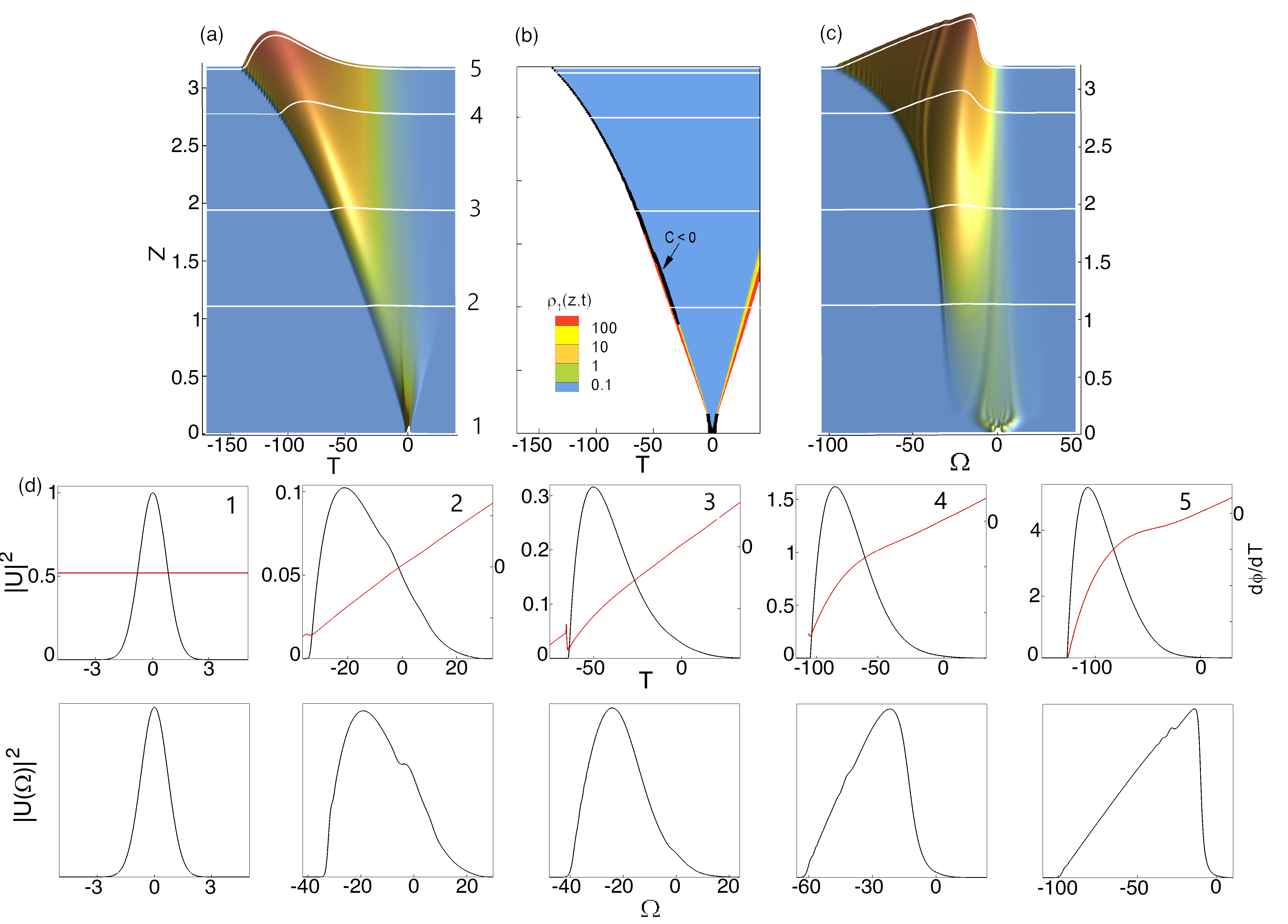}
\caption{Formation of asymmetric nonlinear pulses without "standard" wave breaking: (a) and (c) show 3d evolution of pulse intensity and power spectral density, respectively.  Figure (b) shows in the plane (Z,T) a function $\rho_{1}(Z,T)= 0.5\,\epsilon\, |U(Z,T)|^{-3} \,\frac{\partial^2 |U(Z,T)|}{\partial T^2}$ that defines applicability of the "quasi-classical" approximation. The middle row shows pulse intensity (black lines) and instantaneous frequency (chirp) at $Z_k$ corresponding to the horizontal lines marked with the index $k$ in the upper row. The bottom row shows the power spectral density  at  $Z_k$. Here $\epsilon = 0.004$, $ \delta = 0.045$.}
\label{fig:evolution_free}
\end{figure}

The simple way to prove this is to make a transformation $A(z,t) = B(z,t) \exp[ -g_1 t/\beta_2]$ with the equation for the field $B$ being conservative. 

In this Letter we consider evolution of the initial Gaussian pulse $A(t,z=0)= \sqrt{P_0} \exp[-t^2/(2 T_0^2)]$ in Eq.~\ref{eq:dimNLSE}. In the linear case ($\gamma=0$) pulse  experiences amplification combined with the conventional broadening, while  continuously accelerating:  position of the pulse $t_p(z)$ peak power is changing following the parabolic trajectory:
$  t_p(z) =-\beta_2 g_1 z^2/T_0^2.$
Direction of the drift is defined by the sign of the product of dispersion and gain slope parameter $\beta_2 g_1$. Gain slope leads to the continuous  shift of the position of the pulse spectral power maximum in the frequency domain: $ \omega_p(z)= -g_1 z/T_0^2.$  In the linear medium evolution of the considered initial pulse preserves its symmetric Gaussian shape  both in time and frequency domains, see Supplementary Material for details.

In the nonlinear regime, however, initially symmetric waveform  evolves into asymmetric pulse. 
It is convenient to re-write Eq. \ref{eq:dimNLSE} in the normalised form with two dimensionless parameters $\epsilon$ and $\delta$: 
\begin{equation}
	i \frac{\partial U }{\partial Z} - \frac{1}{2} \frac{\partial^2 U}{\partial T^2} + \frac{1}{\epsilon}\; |U|^2 U = \delta \,\frac{\partial U }{\partial T},
    \label{eq:dimNLSE1}
\end{equation}
here $A(z,t)=\sqrt{P_0} \, U(Z,T)$, $T=t/T_0$ (and $\Omega = \omega T_0$), $Z=z/L_{dis},$ with $L_{dis}= T_0^2/|\beta_2|.$ We consider the so-called normal dispersion medium with $\beta_2 >0$.  Two dimensionless parameters $\epsilon= |\beta_2|/(\gamma P_0 T_0^2)=L_{NL}/L_{dis}$ (see e.g. \cite{Agrawal2010}) and $\delta =  g_1 T_0/|\beta_2|$ define the pulse evolution. Parameter $\epsilon$ characterizes interplay between dispersive and nonlinear effects and is small in highly nonlinear regimes, that we examine here. 

Similar to amplification with constant gain \cite{PP01,PP001}, depending on the initial power and temporal width (that correspond to different points in the ($\epsilon, \,\delta$) plane), Gaussian pulse evolution in Eq.~\ref{eq:dimNLSE1}  can lead to regimes with: (i) the "standard wave breaking" defined as an overtaking of different parts of the pulse, and nonlinear generation of new frequencies during overtaking, or (ii) without wave breaking, see for details Supplementary. As we will show below, type of the nonlinear evolution  is determined by an interplay between the so-called "quantum pressure" term in Eq.~\ref{eq:dimNLSE1} (see for details e.g. \cite{SW01,SW02}): $\,\frac{1}{2 \,|U|} \, \,\frac{\partial^2 |U|}{ \partial T^2} =\rho_1 |U|^{2}/\epsilon$ and the pulse chirp introduced as: $ - \frac{\partial^2}{\partial T^2} \arg(U)$, with the parameter $\rho_1$ (a ratio between the "quantum pressure" and the nonlinear term $|U|^{2}/\epsilon$ in Eq.~\ref{eq:dimNLSE1}) playing an important role in the defining conditions of a wave breaking.  

Propagation of the initial Gaussian with the parameters $\epsilon = 0.004$ ($L_{dis}/L_{NL}=250$), $ \delta = 0.045$ (corresponding to point "1", in the map described in the Supplementary) leading to the formation of the asymmetric pulse is shown in Fig.~\ref{fig:evolution_free}.
Upper row shows 3d dynamics of: (a)  $I(Z,T)=|U(Z,T)|^2$ and (c) $|U(Z,\Omega)|^2$. Middle figure in the upper row depicts in the plane (Z,T) a function $\rho_{1}(Z,T)= 0.5\,  \epsilon \;I^{-3/2} \,\frac{\partial^2 \sqrt{I}}{\partial T^2}$. In the area where $\rho_1 <0.1$ one can apply the Whitham quasi-classical approach \cite{SW00,SW01}, to derive simplified model  Eqs.~\ref{eq:cut1}-\ref{eq:cut2} below.
The middle row in Fig.\ref{fig:evolution_free} shows at the points $Z=Z_k$ (corresponding to the lines marked with the index $k$ in the upper row figures) intensity of the pulse $I(Z_k,T)$ ($k=1,2,3,4,5$) (black lines) and an instantaneous frequency (red lines). The bottom row shows  $|U(Z_k,\Omega)|^2$ at the same  distances $Z_k$. It is seen that while the energy-containing part of the pulse is moving to the negative $T$ experiencing amplification and shape change, the pulse trailing edge stays in the lossy area.  

\begin{figure}[htb]
\centering\includegraphics[width=0.9\columnwidth]{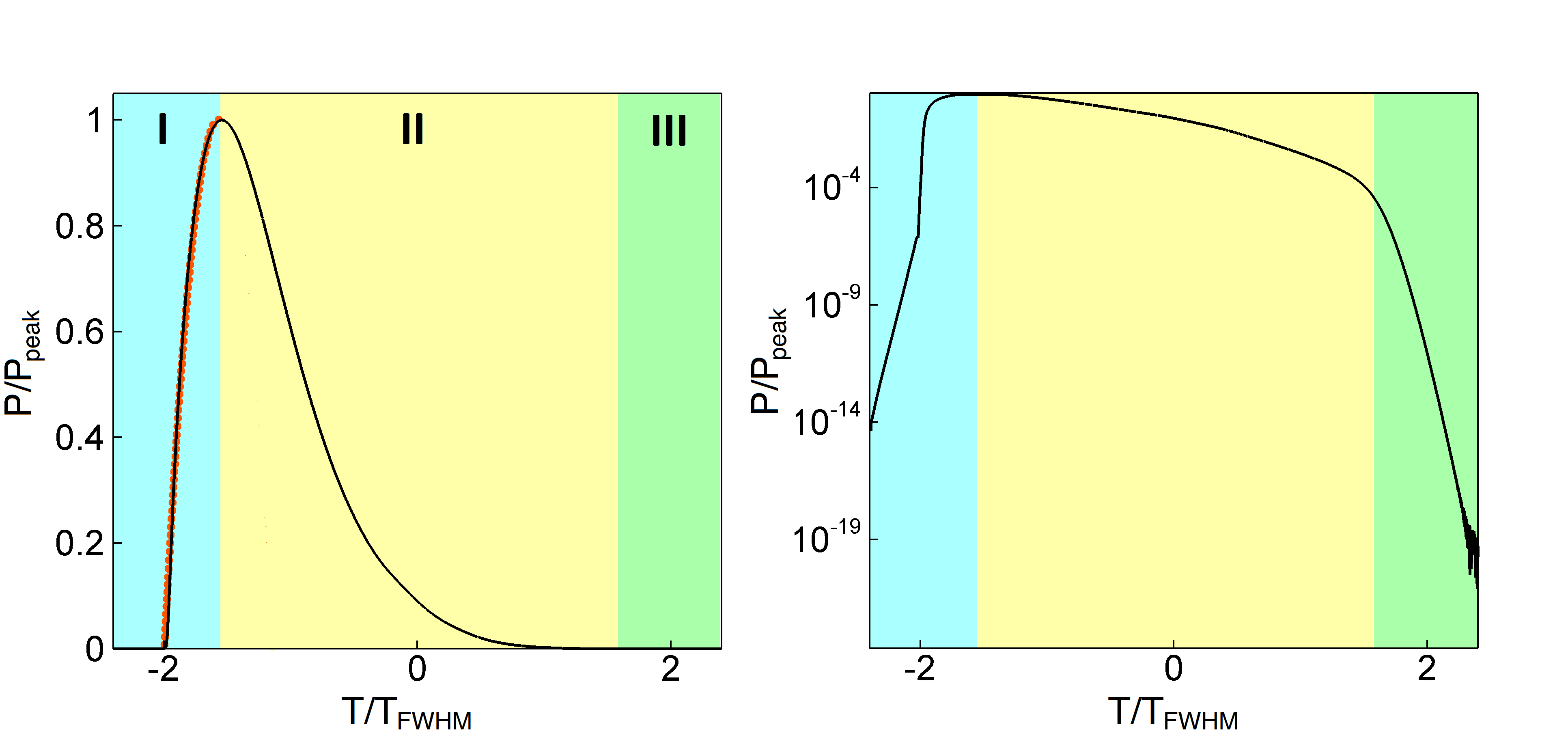}
\caption{Re-scaled details of the temporal shape of a pulse from Fig. 1d-3. Here $|U(Z_3,T)|^2/\max\limits_{T}(|U(Z_3,T)|^2)$ is shown in the normal (left figure) and logarithmic (right figure) scales. $T_{FWHM}$ is the pulse full width at half maximum at $Z=Z_3$.}
%$|U(Z_3,T)|^2/|U(Z_3,0)|^2$
\label{fig:shape}
\end{figure}
Figure \ref{fig:shape} depicts in the linear (left) and logarithmic  (right) scales three characteristic regions across the pulse at the point $Z=Z_3$. Here we change the scales to zoom the pulse structure.  In the zone I,  at the the leading edge, pulse shape is close to parabola similar to the medium with constant gain, in the zone III, pulse has exponential asymptotic determined by the linearised equation.  Parabolic intensity profile at the leading pulse edge allows formation of strongly frequency-swept asymmetric pulse without the degrading effect of the wave breaking \cite{PP02}. In this regime, the change of sign of the chirp ($C=0$), that can be seen in Fig.\ref{fig:evolution_free}d, third figure, occurs in the area where $\rho_1$ is not small and a forming optical shock (gradient catastrophe) is stabilised in the zone I.
 For high pulse powers, when the characteristic nonlinear length $L_{NL}$ is much smaller than the dispersive length $L_{dis}$: $L_{NL}/L_{dis}=\epsilon << 1$, it is customary to consider the so-called quasi-classical limit (see, e.g. \cite{SW00,SW01,SW02,SW03,SW04,SW05} and references therein), when we can neglect time variations of the field amplitude compared to phase time changes.
 
 After applying the well-known Madelung transformation $I=|U|^2$ and $V= - \frac{\partial}{\partial T} \arg(U)$, \cite{SW00,SW01} we get a reduced model for $I$ and $V$, that is a modification of the one-dimensional shallow water equations \cite{SW00,SW01,SW02,SW06}:  
\begin{equation}  
	\frac{\partial I }{\partial Z} =-\frac{\partial IV }{\partial T} - 2 \delta I V,
	    \label{eq:cut1} 
\end{equation}  
\begin{equation}  
	\frac{\partial V }{\partial Z} =-\frac{\partial  }{\partial T} \frac{V^2}{2} -\frac{1}{\epsilon} \frac{\partial I}{\partial T},
	    \label{eq:cut2}
\end{equation}
Here we neglect the time derivatives $S_1 = \frac{1}{2\sqrt{I}}\frac{\partial^2 \sqrt{I}}{\partial T^2}$ and $S_2 = \frac{\delta}{ 2I} \frac{\partial I}{\partial T}$  compared to $I/\epsilon$ (see Supplementary Material for detail). Evolution of the asymmetric pulse can be described by the modified shallow water equations \ref{eq:cut1}-\ref{eq:cut2} ($\rho_1= \epsilon S_1/I$ and $\rho_2= \epsilon S_2/I$ are both small compared to unity) in the blue area shown in Fig. 1b. Evidently, at the pulse edges this simplified description is not valid. This is a direct analogy with self-similar parabolic solution \cite{Anderson:93,PP01} described by the NLSE or NLSE with the spectrally flat constant gain. The second term in r.h.s. of Eq. \ref{eq:cut1} introduces asymmetry in the amplification. As a result, the leading edge of the pulse is amplified, while the trailing edge decays, which leads to asymmetry. 

 The wave breaking (gradient catastrophe) phenomenon in Eq.~\ref{eq:cut2} is determined by the sign of the chirp. When chip becomes negative at some point: $\partial V / \partial T < 0$, wave breaking is inevitable. The gradient catastrophe manifests itself through the formation of a shock wave (vertical jump in the amplitude in the field $V$), which can be stabilized by the terms neglected in derivation of the Eq.~\ref{eq:cut2}. 
A negative chirp can only occur due to nonlinearity, since the linear dispersion leads to a positive chirp. The self-phase-modulation (SMP)-induced frequency chirp is negative for a convex function with $\partial^2 I / \partial T^2 > 0$ and positive for a concave function  with $\partial^2 I / \partial T^2 < 0$. That is why for a parabolic pulse shape  ($\partial^2 I / \partial T^2 < 0$) SMP-induced frequency chirp is positive and no wave-breaking occurs.  
\begin{figure}[tbp]
\centering\includegraphics[width=0.8\columnwidth]{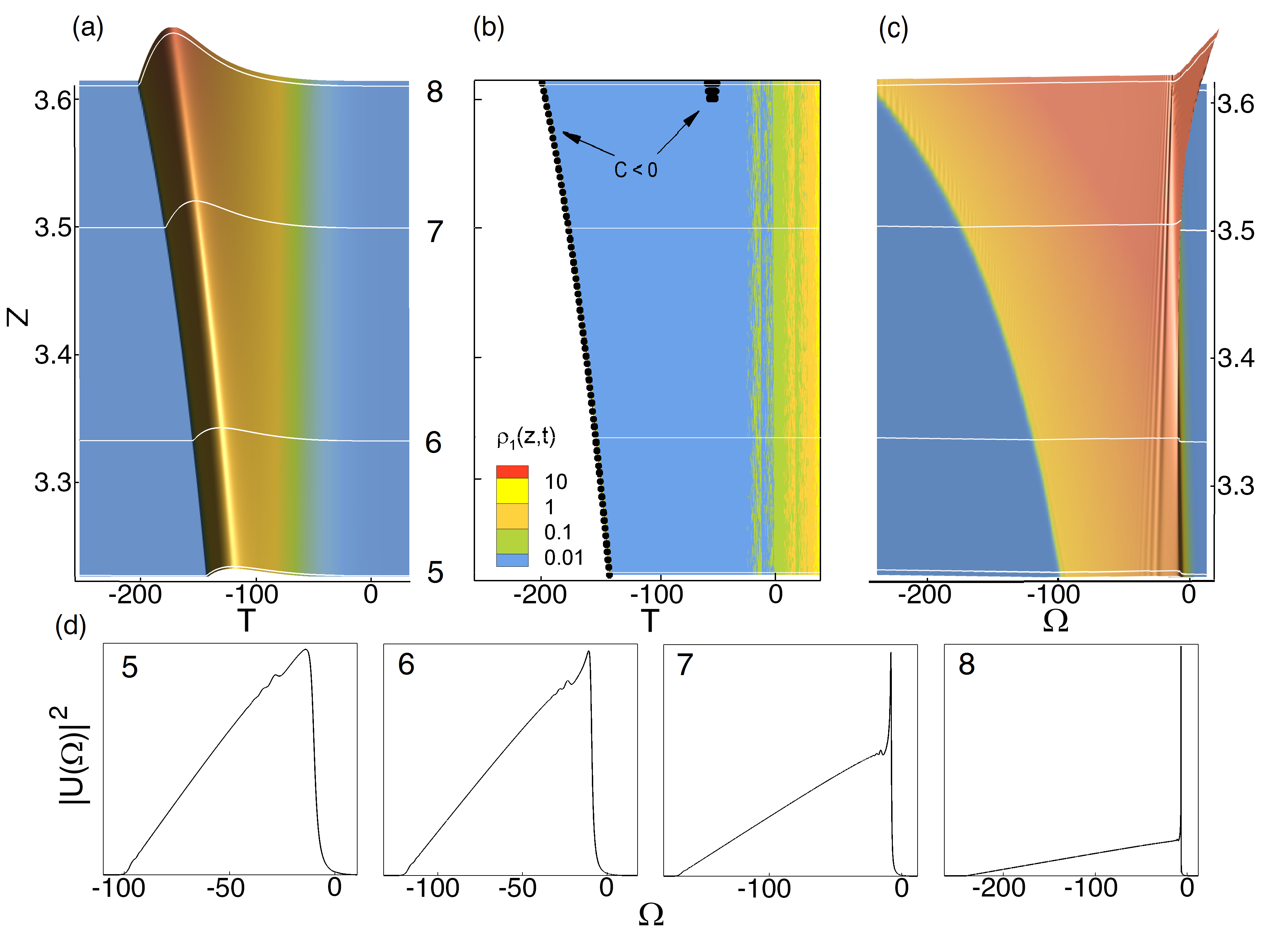}
\caption{Continued Fig.\ref{fig:evolution_free}: (a) and (c) show 3d evolution of pulse intensity and power spectral density, respectively.  Middle figure in the upper row shows a function $\rho_{1}(Z,T)$. The bottom row shows the power spectral density at the points $Z_k$ corresponding to the horizontal lines marked with the index $k$ in the upper row figures. Here $\epsilon = 0.004$, $ \delta = 0.045$.}
\label{fig:break1}
\end{figure}
Figure \ref{fig:evolution_free}b shows the parameter $\rho_{1}(Z,T) $ across the pulse. The black circles indicate an area of negative chirp where the process of wave breaking triggers. It is seen that the chirp is negative at the leading edge of the pulse (with the parabolic shape), where $\rho_{1}$ is not small anymore and the temporal derivative $S_1$ cannot be neglected and this stabilizes wave breaking. 
\begin{figure}[h]
\centering\includegraphics[width=0.7\columnwidth]{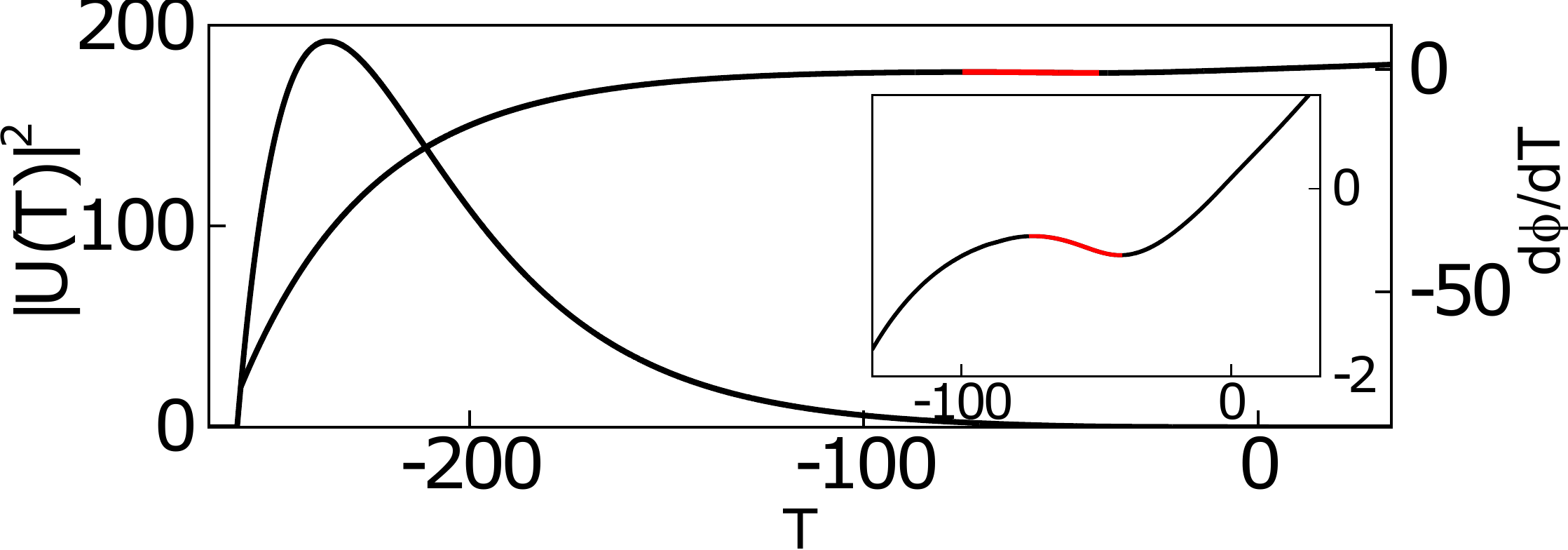}
\caption{Details of the temporal pulse shape and instantaneous frequency of the asymmetric pulse shown in Fig.~\ref{fig:break1} before spectral wave breaking (that happens around Z = 3.8). Inset: zoom of the non-monotonic part of the instant frequency.}
\label{fig:break2}
\end{figure}
%\begin{figure}[h]
%\centering\includegraphics[width=\linewidth]{Figure3_a.png}
%\caption{Wave-breaking free evolution of temporal and spectral pulse shapes.}
%\label{fig:time}
%\end{figure}
The leading edge prevention of a wave breaking is similar to the parabolic pulse formation in the medium with constant, spectrally flat gain \cite{PP01,PP001,PP03}, where stabilisation is observed at both edges of the pulse. However, the considered asymmetric pulse features different dynamics at the leading and trailing edges, and this asymmetry leads to a new type of spectral wave breaking.

Namely, a build-up of the SMP-induced negative chirp at the trailing edge of the asymmetric pulse leads to development of a spectral optical shock wave, with a formation of high spectral peak in the vicinity of a zero frequency, as shown in Fig.~\ref{fig:break1}. Figure~\ref{fig:break1} continues  Fig.~\ref{fig:evolution_free} (all parameters are the same) showing further propagation of the asymmetric pulse, after it is formed.  It is seen that at the propagation distance $Z$ around $3.8$ chirp becomes negative in the area where $\rho_1$ is small and term with $S_1$ cannot stop spectral wave breaking as seen in Fig.~\ref{fig:evolution_free}d.  Corresponding pulse temporal shape and non-monotonic instantaneous frequency, changing its slope from positive to negative (also shown by black areas in Figure~\ref{fig:break1}b), are depicted in more detail in Fig. ~\ref{fig:break2}). 

\begin{figure}[tbp]
\centering\includegraphics[width=0.8\columnwidth]{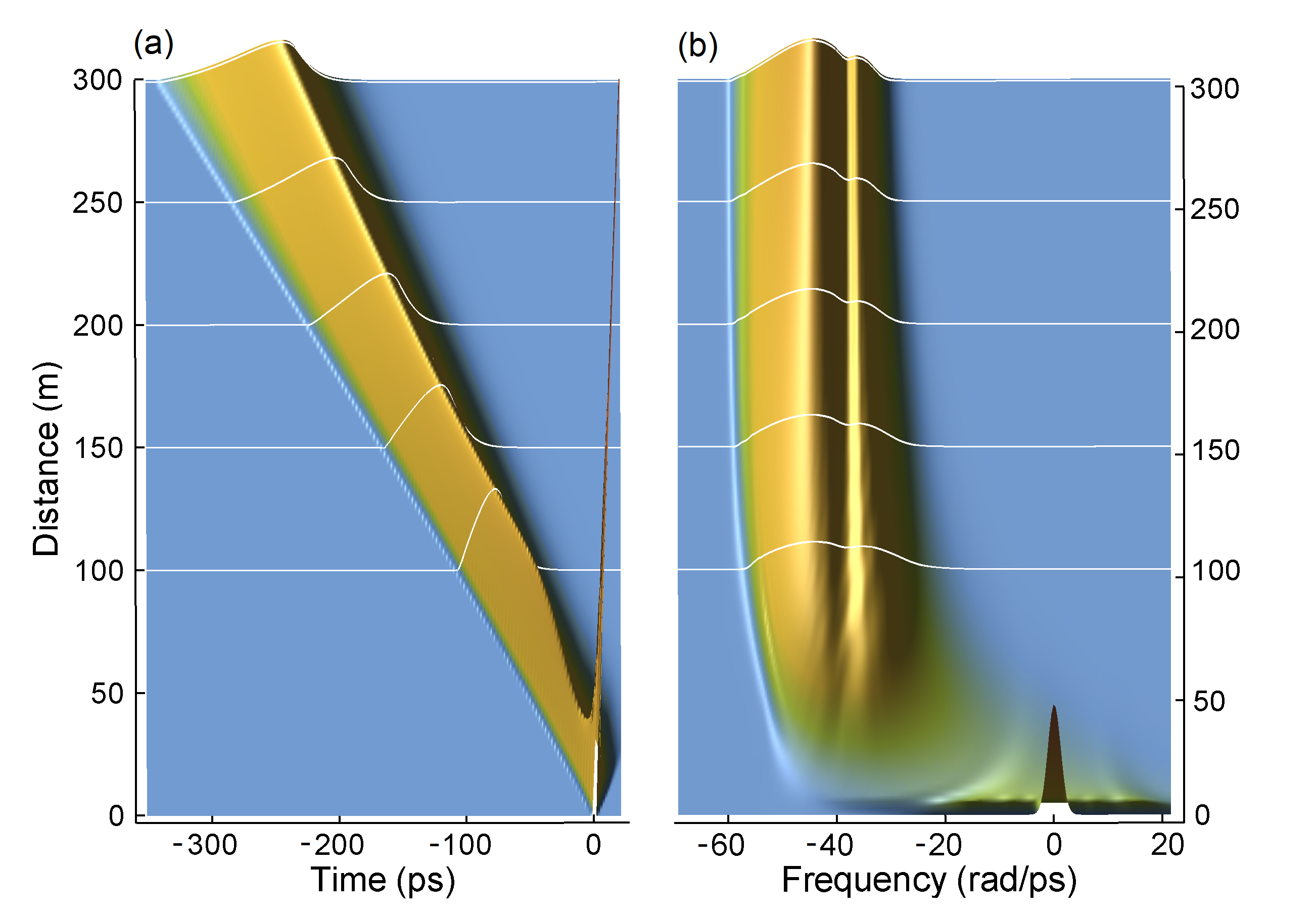}
\caption{Temporal (a) and spectral (b) asymmetric pulse evolution in the case of saturated amplification and finite gain bandwidth.}
\label{fig:time2}
\end{figure}

In more accurate models, the gain is bandwidth-limited, and its slope is linear only within a certain range of frequencies. Additionally, amplification typically becomes saturated at some level.  While the inclusion of these physical effects makes the master model less generic, they combine to stabilize the propagation of nonlinear asymmetric pulses, preventing wave breaking, as shown in Fig.~\ref{fig:time2}. Details of the implementation of the saturation and finite bandwidth of the gain in the numerical modeling are provided in the Supplementary Material. 
%\begin{figure}[tbp]
%\centering\includegraphics[width=0.8\columnwidth]{Figure5_1_spectral_evolution_2D_v2_test.png}
%\caption{Evolution of the spectral pulse shape in logarithmic scale. Formation of spectral peak.}
%\label{fig:break1}
%\end{figure}
As it is seen in Fig.~\ref{fig:time2} asymmetric nonlinear pulse can propagate without wave-breaking, adjusting its spectral and temporal shape to the gain profile.

In conclusion,  a new  model describing optical pulse propagation in a nonlinear dispersive amplifying medium with a gain with linear spectral slope was introduced. An analogy with the hydrodynamic shallow water equations model is discussed. It is shown that negative chirp plays the key role in formation of a spectral optical shock waves in this model system. Observed  asymmetric optical pulses capable to accumulate large nonlinear phase without wave-breaking might be interesting for various applications in high power systems.

The work of A.B. was supported by the Russian Science Foundation (Grant No. 17-72-30006). S.K.T. acknowledges support by the EU project HALT and the Isaac Newton Institute for Mathematical Sciences, Cambridge within the programme HYD2.

%\bibliographystyle{unsrtnat}
%\bibliography{references}  %%% Uncomment this line and comment out the ``thebibliography'' section below to use the external .bib file (using bibtex) .

\section*{Supplemental Materials: Nonlinear optical pulses in media with asymmetric gain}

\section{Linear propagation of a Gaussian pulse in media with a spectral gain slope}

 Linear self-similar evolution of the initial Gaussian pulse $A(t,z=0)= \sqrt{P_0} \exp[-t^2/(2 T_0^2)]$  in the Eq. 1 (with $\gamma=0$ and $g_0=0$) combines amplification and dispersive broadening: 
\begin{equation}
 |A(t,z)|^2=  \frac{T_0 P_0 G(z)}{\tau(z)}  \exp[-\frac{(t-t_p(z))^2}{ \tau^2(z)}], 
\end{equation}
here: 
\begin{equation}
t_p(z) =-\frac{\beta_2 g_1 z^2}{T_0^2},\;\;\tau(z) = T_0 \sqrt{1+\frac{\beta_2^2 \, z^2}{T_0^4}},  \;\;\; G(z)= \exp[ \frac{g_1^2 z^2}{T_0^2}].
\end{equation}
It is seen that the position of the pulse peak power (maximum of $|A(t,z)|^2$ in t) given by $t_{p}(z)$ continuously accelerates with distance $z$ following the parabolic trajectory. Direction of the movement of the pulse peak power is defined by the sign of the product of dispersion $\beta_2$ and the gain slope parameter $g_1$.

The gain slope leads to the continuous  downshift (in the case of a negative gain slope, $g_1>0$) of the position of the maximum of the pulse spectral power in the frequency domain
(here $\omega$ is a detuning from the central frequency $\omega_0$):
\begin{equation}
 |A(\omega,z)|^2 = \frac{T_0^2 P_0}{2 \pi}\, G(z)\, \exp[- T_0^2 \,(\omega - \omega_p)^2], \;\;\; \omega_p= -\frac{g_1 z}{T_0^2}.
 \end{equation}
This spectral shift can be both towards the blue and red regions of the spectrum, and it is a purely linear effect, in contrast to the soliton Raman spectral shift. Potentially, this effect can enable controllable continuous frequency conversion.

Additionally, in a linear medium ($\gamma=0$), the evolution of the initial Gaussian pulse conserves its symmetry and shape in both the time and frequency domains.

We would like to stress once more, that this simplistic "minimal model" for the gain is applied only in the spectral interval that corresponds to the edge of the gain curve approximated by the linear slope. With the field moving outside of this region, other effects should be taken into account that limit and stabilise the energy growth.

\section{Nonlinear propagation with and without "conventional" wave breaking}

\begin{figure}[h]
\centering\includegraphics[width=0.7\columnwidth]{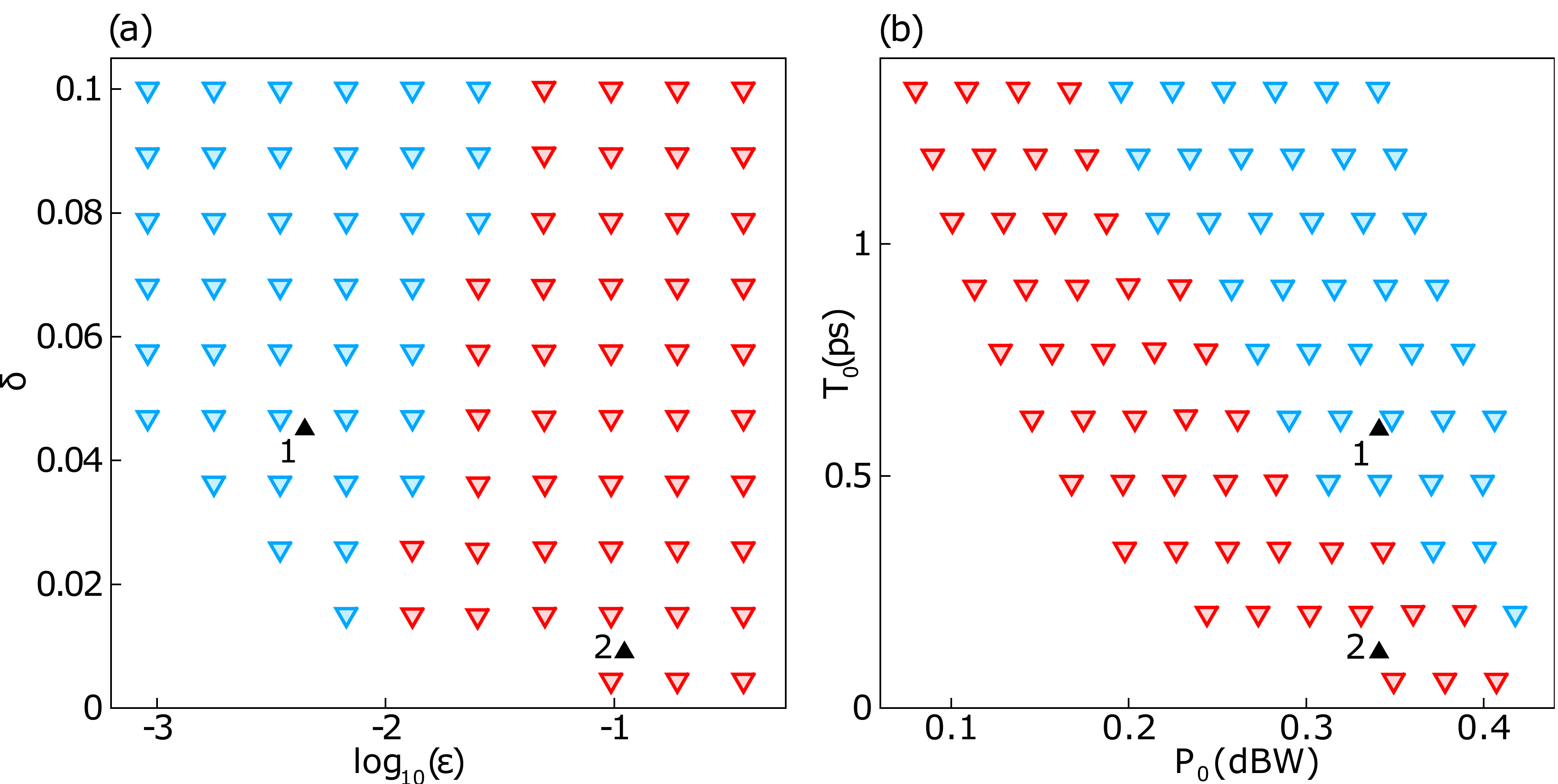}
\caption{Map of the initial Gaussian pulse amplification regimes governed by the Eq. 2 of the main text, with (red triangles) and without (blue triangles) wave breaking in the field of dimensionless parameters $(\epsilon,\delta)$ (a) and dimensional pulse parameters $(P_0\, [W],T_0 \,[ps])$, with x-axis showing power in dBW. Dimensional parameters were calculated for a fixed gain slope $g_1 = 1.5$ ps/km.}
\label{fig:map}
\end{figure}

\begin{figure}[h]
\centering\includegraphics[width=0.9\textwidth]{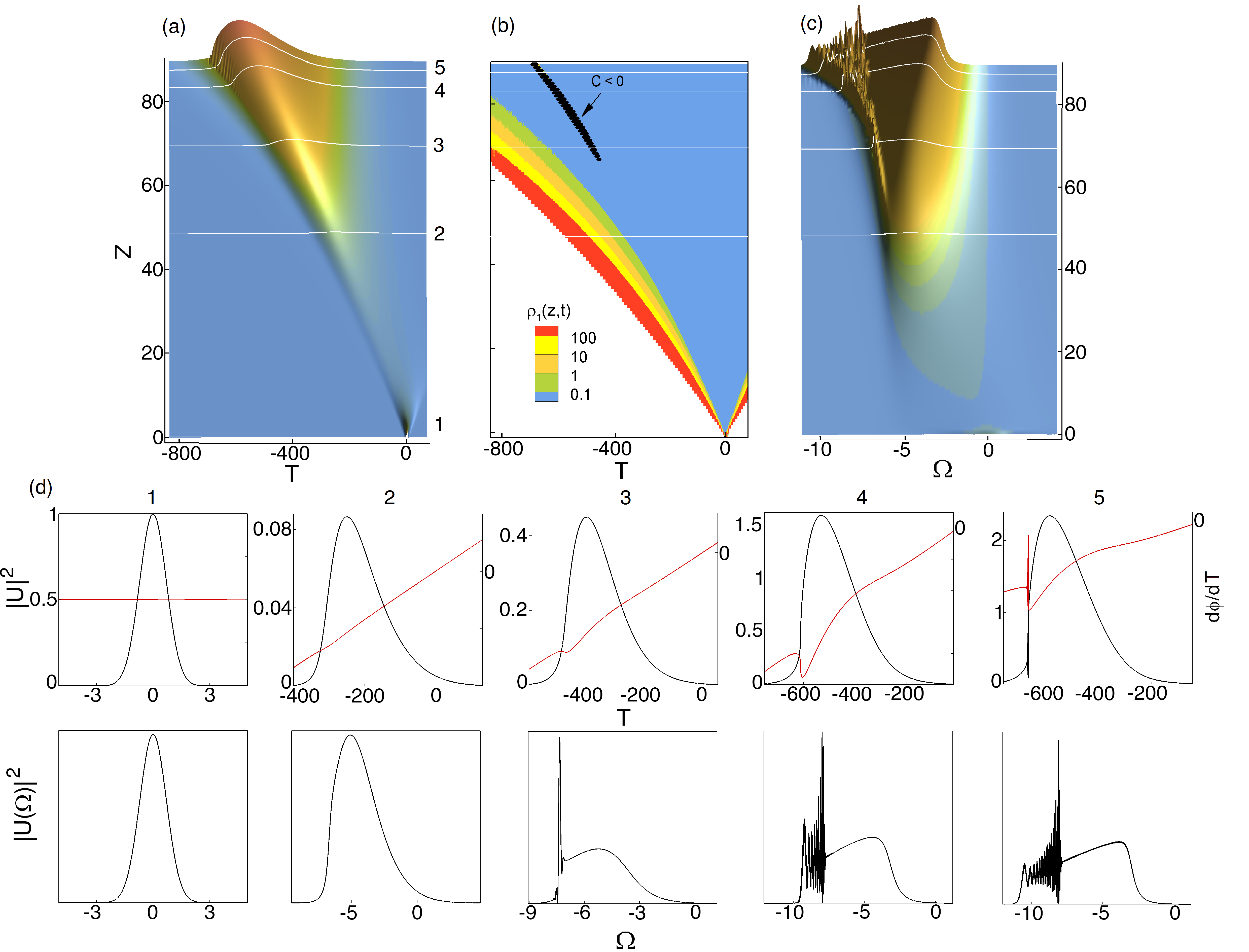}
\caption{Formation of asymmetric nonlinear pulses affected by wave-breaking. Upper row: (a) and (c) 3d evolution of pulse intensity and power spectral density, respectively.  Middle figure in the upper row shows in the plane (Z,T) a function $\rho_{1}(Z,T)= 0.5 \; \epsilon \;  I^{-3/2} \,\frac{\partial^2 \sqrt{I}}{\partial T^2}$. The middle row shows projections of the intensity (black lines) and instantaneous frequency (chirp) at the points $Z_k$ corresponding to the horizontal lines marked with the index $k$ in the upper row figures. The bottom row depicts the spectral power density at the same propagation distances $Z_k$. Here $\epsilon = 0.1$, $ \delta = 0.01$.}
\label{fig:wave_break}
\end{figure}
Conventional wave breaking (also called a gradient catastrophe) is characterised by the two actions: (i) overtaking of different parts of the pulse in time domain and (ii) nonlinear generation of new frequencies during overtaking \cite{Anderson:92,Anderson:93,Wabnitz_2013,SW02}.
We distinguish it from the spectral wave breaking that is illustrated by Fig. 3 of the main text.
Figure \ref{fig:map} shows a map of the wave breaking dynamics with classification of pulse amplification regimes in the plain of dimensionless parameters $\epsilon$ and $\delta$. 
The traditional optical wave breaking is a result of the interplay between dispersion and the self-phase-modulation (SPM) when the pulse propagates  in the normal dispersion regime.
The SMP-induced frequency chirp produces the red-shifted light
near the leading edge traveling faster then the central part of the pulse, and the blue-shifted field traveling slower (than
the  not shifted pulse parts)  near the trailing-edge.

The criterion for wave breaking is a non-monotonically varying pulse intensity in the energy-containing part of the pulse: $\partial |U(Z,T)|^2 / \partial T = 0$ for $|U(Z,T)|^2/\max\limits_{T}(|U(Z,T)|^2) > 10^{-2}$ \cite{Anderson:92}, which signals the presence of characteristic oscillations caused by generation of different frequencies during the overtaking. We consider in detail two points on the plain, marked by "1" and "2". Wave-breaking-free propagation occurs in point "1", when nonlinear length is much smaller than the dispersion length ($\epsilon \ll 1$) for initial Gaussian pulse. This case is discussed in the main text, see Fig. 1. When the dispersion length becomes comparable with the nonlinear length (point "2") the wave breaks. We would like to reiterate once more than wave breaking free here means that the gradient catastrophe does not happen during the nonlinear formation of the asymmetric pulse at the considered propagation distance (Z = 100). What we call here
wave-breaking free means only that no traditional wave breaking takes place, however, as seen in Fig 3 of the main text, due to the asymmetric amplification, a spectral wave breaking can take place even in the regimes marked by blue.

Here we give more details of the evolution of a Gaussian pulse (point "2", $\epsilon = 0.1$, $\delta = 0.01$) affected by the conventional wave breaking (Fig.~\ref{fig:wave_break}). The steepening of the instantaneous frequency occurs because of high nonlinear phase shift, leading to the negative chirp. Nonlinearity affected the pulse evolution before the parabolic shape is formed. As a result, the red-shifted light near the leading edge of the pulse outrun the unshifted light in the forward tail of the pulse, leading to optical wave-breaking \cite{Tomlinson:85,Anderson:92}. Overtaking of different pulse components on the leading edge of the temporal intensity profile  results in rapid oscillations in the spectral profile, which are clearly seen in Fig.~\ref{fig:wave_break}d. It is seen in Fig.~\ref{fig:wave_break}b that chirp is negative in the area when $\rho_1$ is small. This  means, that the temporal derivatives cannot smooth the oscillations. In the absence of stabilizing mechanism (as in the parabolic pulse) the oscillations characterising wave breaking occur  both in the full model Eq. (2) and in the simplified  Eqs. (3-4) discussed in the main text. 

We would like to re-iterate that Fig. 3 in the main text demonstrates that the propagation regimes that do not feature "traditional" wave breaking (the gradient catastrophe in time domain) with further evolution might result into the spectral wave breaking as it is seen in Fig. 3d in the main text. 

\section{The Madelung transformation and modified shallow water equations}
The Madelung transform \cite{madelung1927quantentheorie} is known to relate Schrödinger-type equations (in our case the nonlinear Schrödinger equation) and the hydrodynamic equations. Substitution of 
\[  U(Z,T) = \sqrt{I(Z,T)}\ exp\left[-i\phi(Z,T)\right]\]
into the Eq. (2) in the main text yields:
\begin{equation}
\frac{\partial I}{\partial Z}  = -\frac{\partial }{\partial T} I\phi_T - 2\delta\frac{\partial \phi}{\partial T} I 
\end{equation}
\begin{equation}
\frac{\partial \phi}{\partial Z}  = \frac{1}{2\sqrt{I}}\frac{\partial^2 \sqrt{I}}{\partial T^2}  + \frac{\delta}{2I}\frac{\partial I}{\partial T} - \frac{1}{2} \phi_T^2 - \frac{1}{\epsilon} I = S_1(z,t) + S_2(z,t) + S_3(z,t) + S_4(z,t) 
\end{equation}

%\begin{gather}
%\frac{\partial I}{\partial z}  = \frac{\partial }{\partial t} I\phi_t + 2g_1\frac{\partial \phi}{\partial t} I = R_1(z,t) + R_2(z,t) \\
%\frac{\partial \phi}{\partial z}  = -\frac{1}{2\sqrt{I}}\frac{\partial^2 \sqrt{I}}{\partial t^2} + \frac{1}{2} \phi_t^2 + I - \frac{g_1}{2} \frac{\partial ln(I)}{\partial t} = S_1(z,t) + S_2(z,t) + S_3(z,t) + S_4(z,t) 
%\end{gather}

To evaluate relative contributions and importance of the terms in the r.h.s. of Eq. S4, let us defined  functions (here, evidently, $|\rho_4|=1$):
\begin{equation}
\rho_i(Z,T) = \epsilon \; \frac{S_i(Z,T)}{I(Z,T)},\ i=1,2,3
\end{equation}

Analysis of the evolution of initial Gaussian pulse with different parameters corresponding to different areas in the plane $\epsilon$ and $\delta$ shows that that in the energy-containing parts of the pulse we can simplify description by neglecting $S_1$ and $S_2$ compared to $S_4$ (or in other terms, assuming $\rho_1(Z,T) \ll 1$  and  $\rho_2(Z,T)\ll 1$) in Eq. S4. 
This is similar to applying traditional parabolic pulse approximation in the NLSE \cite{Anderson:93} or in the NLSE with constant (spectrally symmetric) gain \cite{PP01}. 
For instance, Fig. \ref{fig:wave_break}b shows the area (blue) where $\rho_1$ is small. In this area $\rho_2$ also can be neglected.
This leads to a simplified set of equations in the following form:

\begin{equation}
\frac{\partial I}{\partial Z}  = -\frac{\partial }{\partial T} I\phi_T - 2\delta \frac{\partial \phi}{\partial T} I
\end{equation}
\begin{equation}
\frac{\partial \phi}{\partial Z}  =  - \frac{1}{2} \phi_T^2 - \frac{1}{\epsilon}  I 
\end{equation}

Note that when $\delta=0$ this is the well-known shallow water equations (see e.g. \cite{SW02} and references therein).
Introducing an analog of the hydrodynamical velocity:
\begin{equation}
\frac{\partial \phi}{\partial T} = V 
\end{equation}
one can re-write equations in a more traditional hydrodynamic format:
\begin{equation}
	\frac{\partial I }{\partial Z} =-\frac{\partial IV }{\partial T}  - 2\delta I V, 
\end{equation}
\begin{equation}
	\frac{\partial V }{\partial Z} =-\frac{\partial  }{\partial T} \frac{V^2}{2} - \frac{1}{\epsilon} \frac{\partial I}{\partial T},
\end{equation}
This is a novel dissipative modification of the well known shallow water equations \cite{SW02}. 

\section{Stabilisation of nonlinear asymmetric pulses}

\begin{figure*}[t]
\centering\includegraphics[width=0.4\textwidth]{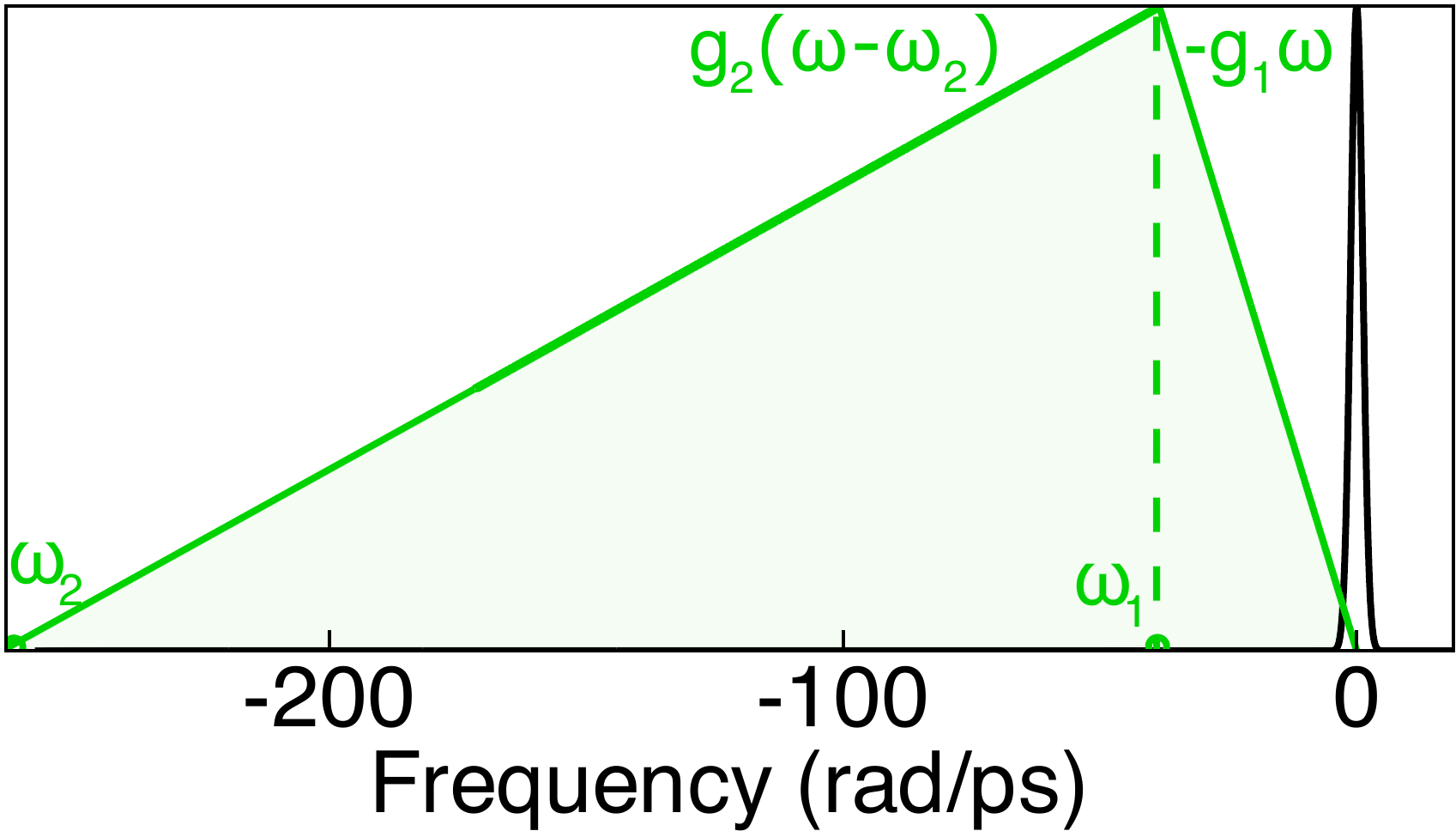}
\caption{Gain profile $g_{FB}(\omega)$ (solid green line) and spectral shape of the initial Gaussian pulse (black line).}
\label{fig:gain}
\end{figure*}

Though model described by Eq. 1 introduced in the main text is generic, it evidently misses important physical effects such as finite bandwidth of the gain and saturation of the amplification. Equation \ref{eq:dimNLSE2} includes two terms in the r.h.s that provide simple generalisation of Eq. 1 in the main text accounting for a gain saturation (term with $\kappa$) and finite bandwidth (first term in r.h.s.):

\begin{equation}
	i \frac{\partial A }{\partial z} - \frac{\beta_2}{2} \frac{\partial^2 A}{\partial t^2} + \gamma |A|^2 A = i \int_{-\infty}^{\infty}g_{FB}(\omega)\tilde A(z,\omega) e^{-i\omega t} d\omega - i \kappa \left(\int_{-\infty}^{\infty}{|A(z,t)|^2dt} \right) A
    \label{eq:dimNLSE2}
\end{equation}

Here $g_{FB}$ is the spectral gain profile (see Fig.~\ref{fig:gain}) given by the following expression (note that $\omega_1$ and $\omega_2$ here are negative):
$$
g_{FB}(\omega) = \begin{cases} 
-g_1 \omega, & \text{if }  \omega > \omega_1 \\ g_{max}(\omega - \omega_2)/(\omega_1 - \omega_2), & \text{if }  \omega < \omega_1. \end{cases}
$$
here the gain slope $g_2$ was added to make the amplification bandwidth finite, frequencies $\omega_1$ and $\omega_2$ correspond to the gain maximum and the left end of the gain curve, respectively. We use the following parameters in the numerical modeling: $g_1 = 1.5$ ps/km, $g_{max} = -g_1 \omega_1$, $\omega_1 = -38$ rad/ps, $\omega_2 = -263$ rad/ps, and the coefficient of the gain saturation $\kappa=0.01$.
Equation reflects the key physical effects accurately studied in \cite{Sidorenko:19}, while still presenting them in a simplified form that can be used in different physical systems.

\begin{figure*}[h]
\centering\includegraphics[width=0.7\textwidth]{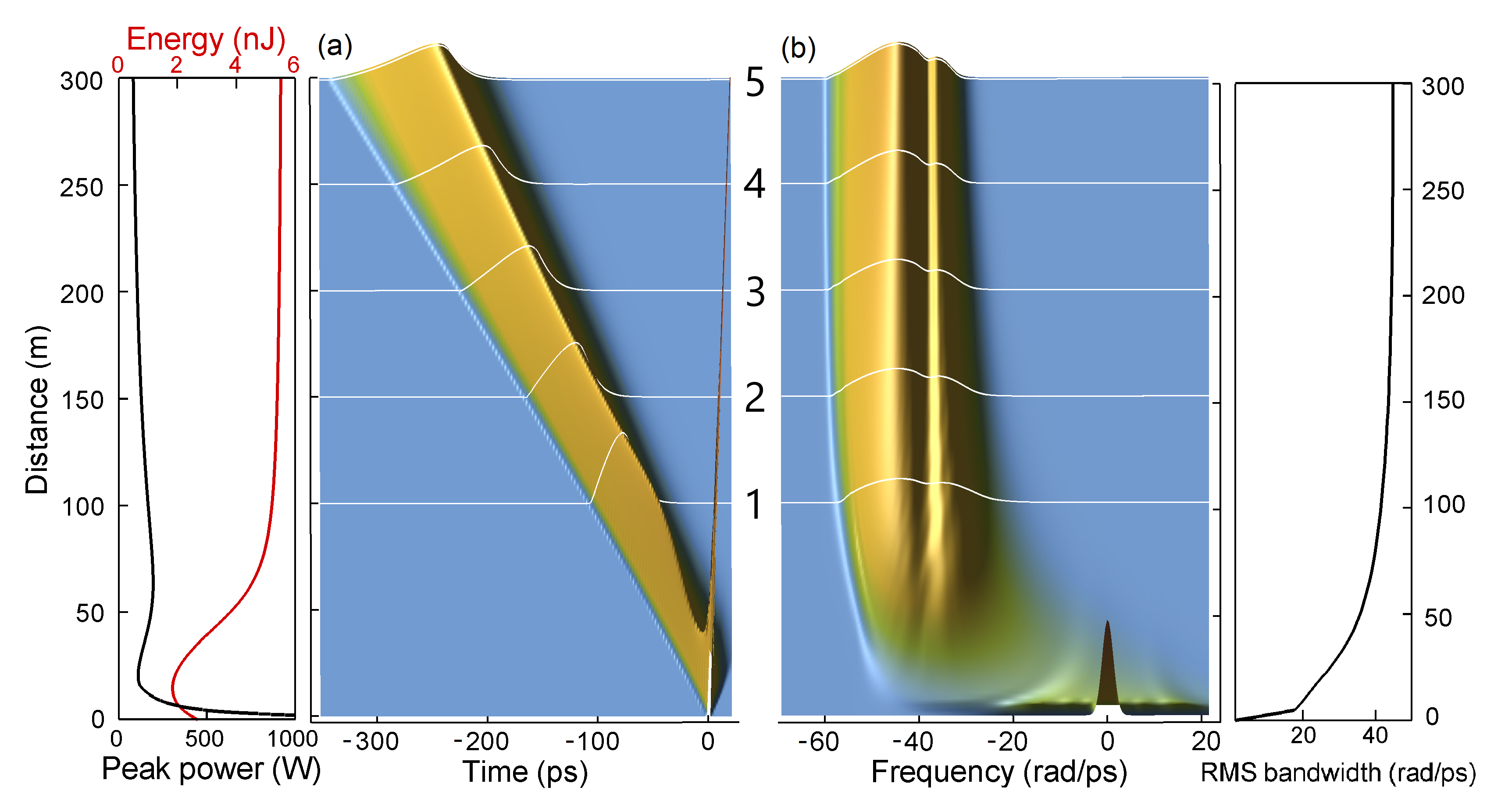}
\caption{Evolution of temporal (a) and spectral (b) pulse shapes in presence of gain saturation and finite gain spectrum bandwidth. Left inset shows evolution of the peak power and energy. Right inset shows dynamics of the RMS pulse bandwidth.}
\label{fig:time}
\end{figure*}

\begin{figure*}[h]
\centering\includegraphics[width=0.8\textwidth]{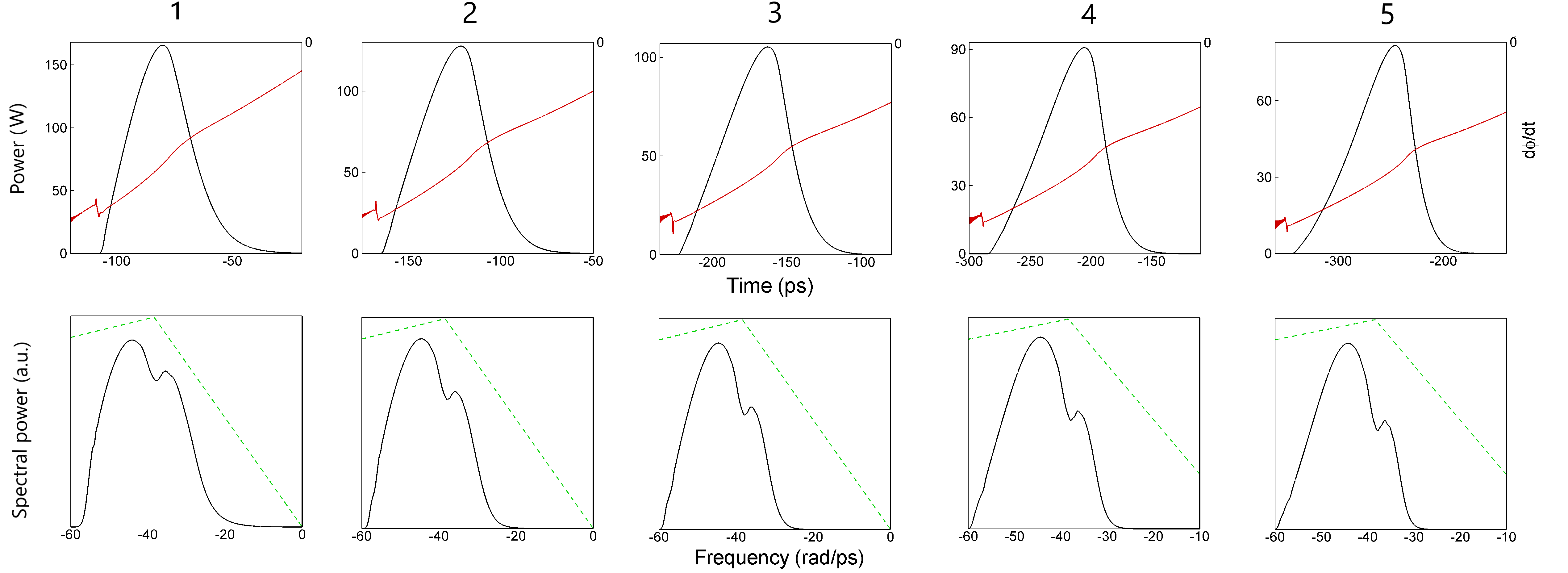}
\caption{Temporal (upper row) and spectral (bottom row) pulse shapes at the points $Z_k$ corresponding to the horizontal lines marked with the index k in the Fig.~\ref{fig:time}. }
\label{fig:time1}
\end{figure*}

 Figure \ref{fig:time} depicts temporal (a) and spectral (b) pulse evolution with propagation in the model Eq. \ref{eq:dimNLSE}. Left inset shows evolution of the peak power and energy. Right inset shows dynamics of the RMS (root-mean-square) pulse bandwidth. The parameters of the initial Gaussian pulse correspond to the point "1" in Fig.~\ref{fig:map}: $\epsilon = 0.004$, $ \delta = 0.045$ ($P_{peak} =2.5 $ kW, $T_{FWHM} = 1$ ps in dimension units). 

%\begin{figure*}[h]
%\centering\includegraphics[width=0.6\textwidth]{shape_asymptotics.pdf}
%\caption{Evolution of temporal and spectral pulse shapes in presence %of gain saturation and finite gain spectrum bandwidth}
%\label{fig:time1}
%\end{figure*}

Figure \ref{fig:time1} provide more details on the changes in the pulse intensity and power spectral density with evolution. Upper row shows at the points $z=z_k$ (corresponding to the lines marked with the index $k$ in the Fig.\ref{fig:time}) temporal shape of the pulse $|A(z_k,t)|^2$ ($k=1,2,3,4,5$). Bottom row shows spectral shape of the pulse $|\tilde{A}(z_k,\omega)|^2$ ($k=1,2,3,4,5$). Gain spectrum $g_{FB}(\omega)$ is depicted by the dashed green line. 

\bibliographystyle{unsrtnat}
\bibliography{references}

\providecommand{\noopsort}[1]{}\providecommand{\singleletter}[1]{#1}%
\begin{thebibliography}{23}
\providecommand{\natexlab}[1]{#1}
\providecommand{\url}[1]{\texttt{#1}}
\expandafter\ifx\csname urlstyle\endcsname\relax
  \providecommand{\doi}[1]{doi: #1}\else
  \providecommand{\doi}{doi: \begingroup \urlstyle{rm}\Url}\fi

\bibitem[Siegman(1986)]{siegman1986lasers}
A.~E. Siegman.
\newblock \emph{Lasers}.
\newblock University Science Books, 1986.

\bibitem[Agrawal(2010)]{Agrawal2010}
G.~Agrawal.
\newblock \emph{Applications of Nonlinear Fiber Optics}.
\newblock Elsevier Science, 2010.

\bibitem[Dienes et~al.(1996)Dienes, Heritage, Jasti, and Hong]{Dienes:96}
A.~Dienes, J.~P. Heritage, C.~Jasti, and M.~Y. Hong.
\newblock Femtosecond optical pulse amplification in saturated media.
\newblock \emph{J. Opt. Soc. Am. B}, 13\penalty0 (4):\penalty0 725--734, 1996.

\bibitem[Ruter et~al.(2010)Ruter, Makris, El-Ganainy, Christodoulides, Segev,
  and Kip]{Ruter2010}
C.~E. Ruter, K.~G. Makris, R.~El-Ganainy, D.~Christodoulides, M.~Segev, and
  D.~Kip.
\newblock Observation of parity-time symmetry in optics.
\newblock \emph{Nature Physics}, 6\penalty0 (3):\penalty0 192--195, 2010.

\bibitem[Sidorenko et~al.(2019)Sidorenko, Fu, and Wise]{Sidorenko:19}
P.~Sidorenko, W.~Fu, and F.~Wise.
\newblock Nonlinear ultrafast fiber amplifiers beyond the gain-narrowing limit.
\newblock \emph{Optica}, 6\penalty0 (10):\penalty0 1328--1333, 2019.

\bibitem[Fermann et~al.(2000)Fermann, Kruglov, Thomsen, Dudley, and
  Harvey]{PP01}
M.~Fermann, V.~Kruglov, B.~C. Thomsen, J.~M. Dudley, and J.~D. Harvey.
\newblock Self-similar propagation and amplification of parabolic pulses in
  optical fibers.
\newblock \emph{Phys. Rev. Lett.}, 84:\penalty0 6010--6013, 2000.

\bibitem[Dudley et~al.(2007)Dudley, Finot, Richardson, and Millot]{PP02}
J.~M. Dudley, C.~Finot, D.~J. Richardson, and G.~Millot.
\newblock Self-similarity in ultrafast nonlinear optics.
\newblock \emph{Nature Physics}, 3:\penalty0 597--603, 2007.

\bibitem[Ilday et~al.(2004)Ilday, Buckley, Clark, and Wise]{PP03}
F.~Ilday, J.~R. Buckley, W.~G. Clark, and F.~W. Wise.
\newblock Self-similar evolution of parabolic pulses in a laser.
\newblock \emph{Phys. Rev. Lett.}, 92:\penalty0 213902, 2004.

\bibitem[Boscolo et~al.(2002)Boscolo, Turitsyn, Novokshenov, and Nijhof]{PP04}
S.~Boscolo, S.~K. Turitsyn, V.~Y. Novokshenov, and J.~H.~B. Nijhof.
\newblock Self-similar parabolic optical solitary waves.
\newblock \emph{Theoretical and Mathematical Physics}, 133\penalty0
  (3):\penalty0 1647--1656, 2002.

\bibitem[Finot et~al.(2006)Finot, Parmigiani, Petropoulos, and
  Richardson]{Finot:06}
C.~Finot, F.~Parmigiani, P.~Petropoulos, and D.~J. Richardson.
\newblock Parabolic pulse evolution in normally dispersive fiber amplifiers
  preceding the similariton formation regime.
\newblock \emph{Opt. Express}, 14\penalty0 (8):\penalty0 3161--3170, 2006.

\bibitem[Kruglov et~al.(2000)Kruglov, Peacock, Dudley, and Harvey]{PP001}
V.~I. Kruglov, A.~C. Peacock, J.~M. Dudley, and J.~D. Harvey.
\newblock Self-similar propagation of high-power parabolic pulses in optical
  fiber amplifiers.
\newblock \emph{Opt. Lett.}, 25\penalty0 (24):\penalty0 1753--1755, 2000.

\bibitem[El and Hoefer(2016)]{SW01}
G.~A. El and M.A. Hoefer.
\newblock Dispersive shock waves and modulation theory.
\newblock \emph{Physica D: Nonlinear Phenomena}, 333:\penalty0 11--65, 2016.

\bibitem[Biondini et~al.(2016)Biondini, El, Hoefer, and Miller]{SW02}
G.~Biondini, G.A. El, M.A. Hoefer, and P.D. Miller.
\newblock Dispersive hydrodynamics: Preface.
\newblock \emph{Physica D: Nonlinear Phenomena}, 333:\penalty0 1--5, 2016.

\bibitem[Whitham(2011)]{SW00}
G.~B. Whitham.
\newblock \emph{Linear and Nonlinear Waves}.
\newblock Pure and Applied Mathematics. Wiley, 2011.

\bibitem[El et~al.(2001)El, Grimshaw, and Pavlov]{SW03}
G.~A. El, R.~H.~J. Grimshaw, and M.~V. Pavlov.
\newblock Integrable shallow-water equations and undular bores.
\newblock \emph{Studies in Applied Mathematics}, 106\penalty0 (2):\penalty0
  157--186, 2001.

\bibitem[Xu et~al.(2016)Xu, Garnier, Faccio, Trillo, and Picozzi]{SW04}
G.~Xu, J.~Garnier, D.~Faccio, S.~Trillo, and A.~Picozzi.
\newblock Incoherent shock waves in long-range optical turbulence.
\newblock \emph{Physica D: Nonlinear Phenomena}, 333:\penalty0 310--322, 2016.

\bibitem[Xu et~al.(2017)Xu, Conforti, Kudlinski, Mussot, and Trillo]{SW05}
G.~Xu, M.~Conforti, A.~Kudlinski, A.~Mussot, and S.~Trillo.
\newblock Dispersive dam-break flow of a photon fluid.
\newblock \emph{Phys. Rev. Lett.}, 118:\penalty0 254101, 2017.

\bibitem[Bendahmane et~al.(2022)Bendahmane, Xu, Conforti, Kudlinski, Mussot,
  and Trillo]{SW06}
A.~Bendahmane, G.~Xu, M.~Conforti, A.~Kudlinski, A.~Mussot, and S.~Trillo.
\newblock The piston riemann problem in a photon superfluid.
\newblock \emph{Nature Commun.}, 13\penalty0 (3137), 2022.

\bibitem[Anderson et~al.(1993)Anderson, Desaix, Karlsson, Lisak, and
  Quiroga-Teixeiro]{Anderson:93}
D.~Anderson, M.~Desaix, M.~Karlsson, M.~Lisak, and M.~Quiroga-Teixeiro.
\newblock Wave-breaking-free pulses in nonlinear-optical fibers.
\newblock \emph{J. Opt. Soc. Am. B}, 10\penalty0 (7):\penalty0 1185--1190,
  1993.

\bibitem[Anderson et~al.(1992)Anderson, Desaix, Lisak, and
  Quiroga-Teixeiro]{Anderson:92}
D.~Anderson, M.~Desaix, M.~Lisak, and M.~L. Quiroga-Teixeiro.
\newblock Wave breaking in nonlinear-optical fibers.
\newblock \emph{J. Opt. Soc. Am. B}, 9\penalty0 (8):\penalty0 1358--1361, Aug
  1992.
\newblock \doi{10.1364/JOSAB.9.001358}.
\newblock URL
  \url{https://opg.optica.org/josab/abstract.cfm?URI=josab-9-8-1358}.

\bibitem[Wabnitz(2013)]{Wabnitz_2013}
Stefan Wabnitz.
\newblock Optical tsunamis: shoaling of shallow water rogue waves in nonlinear
  fibers with normal dispersion.
\newblock \emph{Journal of Optics}, 15\penalty0 (6):\penalty0 064002, jun 2013.
\newblock \doi{10.1088/2040-8978/15/6/064002}.
\newblock URL \url{https://dx.doi.org/10.1088/2040-8978/15/6/064002}.

\bibitem[Tomlinson et~al.(1985)Tomlinson, Stolen, and Johnson]{Tomlinson:85}
W.~J. Tomlinson, R.~H. Stolen, and A.~M. Johnson.
\newblock Optical wave breaking of pulses in nonlinear optical fibers.
\newblock \emph{Opt. Lett.}, 10\penalty0 (9):\penalty0 457--459, Sep 1985.
\newblock \doi{10.1364/OL.10.000457}.
\newblock URL \url{https://opg.optica.org/ol/abstract.cfm?URI=ol-10-9-457}.

\bibitem[Madelung(1927)]{madelung1927quantentheorie}
E.~Madelung.
\newblock Quantentheorie in hydrodynamischer form.
\newblock \emph{Zeitschrift f{\"u}r Physik}, 40:\penalty0 322, 1927.

\end{thebibliography}

\end{document}